\documentclass[twocolumn,Conference]{IEEEtran}

\IEEEoverridecommandlockouts
\usepackage[letterpaper, top=0.72in, bottom=0.98in, left=0.68in, right=0.68in]{geometry}

\usepackage{bm}
\usepackage{color}
\usepackage{graphicx}
\usepackage{amsmath}
\usepackage{amssymb}
\usepackage{algorithm}
\usepackage{algorithmic}
\usepackage{ulem}
\usepackage{lineno}
\usepackage{lettrine}
\usepackage{subfigure}
\usepackage{epstopdf}
\usepackage{stfloats}
\usepackage{multirow}
\usepackage{booktabs}
\usepackage{array}
\usepackage{amsthm}
\usepackage{caption}
\usepackage{subfigure}
\usepackage{setspace}
\usepackage{makecell}
\allowdisplaybreaks[4]

\newcommand{\be}{\begin{equation}}
	\newcommand{\ee}{\end{equation}}
\newcommand{\bea}{\begin{eqnarray}}
	\newcommand{\eea}{\end{eqnarray}}
\newcommand{\ba}{\begin{array}}
	\newcommand{\ea}{\end{array}}

\def\BibTeX{{\rm B\kern-.05em{\sc i\kern-.025em b}\kern-.08em
    T\kern-.1667em\lower.7ex\hbox{E}\kern-.125emX}}
\usepackage{balance}

\title{Cooperative Cell-Free ISAC Networks: Joint BS Mode Selection and Beamforming Design 
\thanks{$^{\ast}$ Corresponding author.}
\thanks{This work is supported in part by the National Natural Science Foundation of China (Grant No. 62371090 and 62071083), in part by Liaoning Applied Basic Research Program (Grant No. 2023JH2/101300201), and in part by Dalian Science and Technology Innovation Project (Grant No. 2022JJ12GX014).}
}

\author{\IEEEauthorblockN{Sifan Liu$^{\dag}$, Rang Liu$^{\ddag}$, Zhiping Lu$^{\S}$, Ming Li$^{\dag \ast}$, and Qian Liu$^{\dag}$
		 }
	
\IEEEauthorblockA{$^{\dag}$ Dalian University of Technology, Dalian, Liaoning 116024, China}\\
\IEEEauthorblockA{$^{\ddag}$ University of California, Irvine, CA 92697, USA}\\
\IEEEauthorblockA{$^{\S}$ Beijing University of Posts and Telecommunications, Beijing 100876, China  \\ State Key Laboratory of Wireless Mobile Communications (CICT), Beijing 100191, China }	\\
\IEEEauthorblockA{E-mail: \texttt{sifanliu@mail.dlut.edu.cn, \texttt{rangl2@uci.edu}, \texttt{luzhp$\_$007@163.com}, \{mli,qianliu\}@dlut.edu.cn} }	
\vspace{-0.8cm}}
\begin{document}
\maketitle
\pagestyle{empty}
\thispagestyle{empty}
\begin{abstract}
Owing to the promising ability of saving hardware cost and spectrum resources, integrated sensing and communication (ISAC) is regarded as a revolutionary technology for future sixth-generation (6G) networks.
The mono-static ISAC systems considered in most of existing works can only achieve limited sensing performance due to the single observation angle and easily blocked transmission links, which motivates researchers to investigate cooperative ISAC networks.
In order to further improve the degrees of freedom (DoFs) of cooperative ISAC networks, the transmitter-receiver selection, i.e., base station (BS) mode selection problem, is meaningful to be studied.
However, to our best knowledge, this crucial problem has not been extensively studied in existing works.
In this paper, we consider the joint BS mode selection, transmit beamforming, and receive filter designs for cooperative cell-free ISAC networks, where multi-BSs cooperatively serve communication users and detect targets.
An efficient joint beamforming design algorithm and three different heuristic BS mode selection methods are proposed to solve the non-convex NP-hard problem.
Simulation results demonstrates the advantages of cooperative ISAC networks, the importance of BS mode selection, and the effectiveness of proposed algorithms.
\end{abstract}	
\vspace{-0.3cm}
\vspace{-0.7cm}
\section{Introduction}
Benefiting from the spectrum sharing of communication and sensing, integrated sensing and communication (ISAC) is deemed as a revolutionary technology to solve the spectrum scarcity problem for future sixth-generation (6G) networks \cite{F. Liu1}, \cite{F. Liu2}.
The appealing advantages of ISAC have already triggered many research efforts in mono-static ISAC systems, where single base station (BS) is utilized for both communication and sensing \cite{Z. Xiao}-\cite{Q. Zhu}. 

Unfortunately, due to the single observation angle and complex signal propagation environment easily to be blocked, it is difficult to obtain higher sensing precision or solve complicated sensing problems in such mono-static ISAC system with only one BS. 
These limitations of mono-static systems motivate researchers to investigate multi-static ISAC networks which can provide multi-angle observations and higher spatial diversity \cite{Q. Shi}-\cite{P. Gao}.
Specifically, the trilateration-based target localization problem was considered in \cite{Q. Shi} for multi-cell networks.
The authors in \cite{M. L. Rahman} focused on sensing parameter estimation in their proposed perceptive mobile network (PMN).
The power allocation problem was investigated in \cite{Y. Huang} for networked ISAC.
Moreover, the authors in \cite{P. Gao} considered the Pareto optimization problem in rate-splitting multiple access cooperative ISAC network.

The works introduced above only considered networks where the transmit BSs and receive BSs are fixed.
However, in order to better utilize the additional degrees of freedom (DoFs) provided by multiple BSs and more efficiently use the network resources, the transmit BS and receive BS selection, i.e., the BS mode selection, is also a meaningful problem worth to being studied.
To be specific, richer sensing information and greater spatial diversity gain can be obtained via rational BS mode selection, which provide significant potential to communication and sensing performance improvement \cite{K. Ji}.
Nevertheless, to our best knowledge, this promising problem has not been widely studied in existing works.

Motivated by above discussions, we consider the joint design of BS mode selection, transmit beamforming, and receive filter for a cooperative cell-free ISAC network, where multiple BSs, each of which can operate as either a transmitter or receiver, cooperatively serve communication users and detect targets.
To be specific, we aim to maximize the sum of sensing signal-to-interference-plus-noise ratio (SINR) under the communication SINR requirements, the power budget of the whole network, and the constraints on the numbers of transmit/receive BSs.
In order to solve this non-convex NP-hard problem, we first convert the problem into two sub-problems, i.e., the joint transmit beamforming and receive filter design sub-problem as well as the BS mode selection sub-problem.
Then, an efficient fractional programming (FP) and majorization-minimization (MM) based algorithm is utilized to solve the joint transmit beamforming and receive filter design sub-problem, while three low-complexity heuristic methods are proposed to select the appropriate mode of BSs.
Simulation results verify the importance of BS mode selection in cooperative cell-free ISAC networks and the effectiveness of our proposed joint design algorithms.

\section{System Model and Problem Formulation}

\subsection{System Model}

\begin{figure}[!t]
	\centering
	\includegraphics[width = 2.5 in]{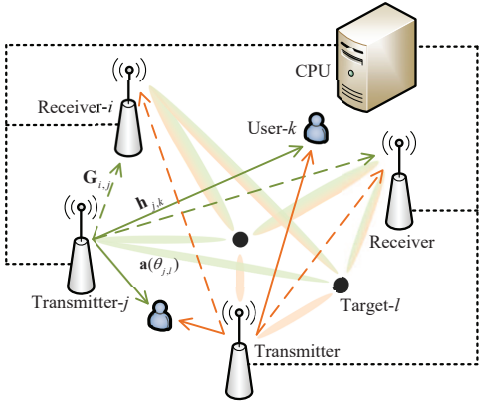}
	\caption{Cooperative cell-free ISAC network.}
	\label{fig:system}
\vspace{-0.4cm}
\end{figure}

We consider a cooperative cell-free ISAC network consists of $J$ BSs, $K$ single-antenna users, and $L$ point-like targets.
Each BS is equipped with $M$ uniform linear array (ULA) antennas and connected to the central processing unit (CPU) for joint transmission and signal processing.
Each BS can work in two modes, i.e., as transmitter or as receiver, which are determined by a functionality selection module and represented by variable $\alpha_j \in \{0,1\},\forall j$.
To be specific, $\alpha_j = 1$ denotes that BS-$j$ operates in the transmitter mode; $\alpha_j = 0$ represents that BS-$j$ operates in the receiver mode.
All the transmit BSs cooperatively serve the communication users and send sensing signals towards the targets.
Meanwhile, the receive BSs collect the echo signals and jointly detect potential targets.

Denote $\mathbf{s}_{\text{c}}\in \mathbb{C}^{{K} \times 1}$ as the communication symbols transmitted to $K$ users, which satisfy $\mathbb{E}\{\mathbf{s}_{\text{c}}\mathbf{s}_{\text{c}}^H\}=\mathbf{I}_{K}$.
Similarly, define $\mathbf{s}_{\text{r}}\in \mathbb{C}^{{M} \times 1}$ as the radar waveforms for targets detection satisfying $\mathbb{E}\{\mathbf{s}_{\text{r}}\mathbf{s}_{\text{r}}^H\}=\mathbf{I}_{M}$.
Assume that $\mathbf{s}_{\text{c}}$ and $\mathbf{s}_{\text{r}}$ are independent, i.e., $\mathbb{E}\{\mathbf{s}_{\text{c}}\mathbf{s}_{\text{r}}^H\}=\mathbf{0}$.
Moreover, we denote the corresponding communication and sensing beamforming matrices from transmit BS-$j$ as $\mathbf{W}_{\text{c},j}\in \mathbb{C}^{M \times K}$ and $\mathbf{W}_{\text{r},j}\in \mathbb{C}^{M \times M}$, respectively.
The transmitted signal from BS-$j$ can be expressed as
\begin{equation}
\mathbf{x}_j = \alpha_j\mathbf{W}_{\text{c},j}\mathbf{s}_\text{c}+\alpha_j\mathbf{W}_{\text{r},j}\mathbf{s}_{\text{r} }= \alpha_j\mathbf{W}_{j}\mathbf{s},
\end{equation}
where $\mathbf{W}_{j} \triangleq [\mathbf{W}_{\text{c},j} \; \mathbf{W}_{\text{r},j}]$ and $\mathbf{s}\triangleq[\mathbf{s}_\text{c}^T \; \mathbf{s}_{\text{r}}^T]^T$.

Then, the received signal at user-$k$ is given by
\begin{equation}
\begin{aligned}
y_{\text{c},k} = \sum_{j =1}^J \mathbf{h}_{j,k}^H \mathbf{x}_j +n_{\text{c},k},\\
\end{aligned}
\end{equation}
in which $\mathbf{h}_{j,k} \in \mathbb{C}^{M \times 1}$ and $n_{\text{c},k} \sim \mathcal{CN} (0, \sigma_{\text{c}}^{2})$ represent the channel from BS-$j$ to user-$k$ and complex additive white Gaussian noise (AWGN) at user-$k$, respectively.
Thus, the received SINR of user-$k$ is given by
\begin{equation}
\begin{aligned}
\mathrm{SINR}_{\text{c},k}=\frac{ \Big| \sum_{j =1}^J \alpha_j\mathbf{h}_{j,k}^H \mathbf{w}_{j,k} \Big|^2}{\sum_{i=1,i \ne k}^{K+M}\Big|\sum_{j =1}^J\alpha_j\mathbf{h}_{j,k}^H \mathbf{w}_{j,i} \Big|^2 + \sigma_{\text{c}}^{2}},
\end{aligned}
\end{equation}
where $\mathbf{w}_{j,k}$ denotes the $k$-th column of $\mathbf{W}_{j}$.

Meanwhile, the transmitted signals are reflected by $L$ targets and then the echo signals are collected by the receive BSs.
Based on the high-density deployment of BSs in practical cell-free networks, we take the interference between BSs into account.
By defining the direct channel from transmit BS-$i$ to receive BS-$j$ as $\mathbf{G}_{i,j} \in \mathbb{C}^{M \times M}$ which satisfy $\mathbf{G}_{i,i} = \mathbf{0}_{M \times M}$ since the BSs can only operate as a transmitter or receiver, the received echo signal at BS-$j$ is written as
\begin{equation}
\begin{aligned}
\hspace{-0.25cm}\mathbf{y}_{\text{r},j }\hspace{-0.05cm}= \hspace{-0.05cm}\sum \limits_{i=1}^J  \sum \limits_{l=1}^L\xi_{j,i,l} \mathbf{a} (\theta_{j,l})\mathbf{a}^T (\theta_{i,l})\mathbf{x}_i +  \sum \limits_{i=1}^J\hspace{-0.05cm}\mathbf{G}_{i,j}^T\mathbf{x}_i +  \mathbf{n}_{\text{r},j},
\end{aligned}
\end{equation}
where $\xi_{j,i,l} \sim \mathcal{CN} (0, \sigma_\text{t}^{2})$ and $\mathbf{n}_{\text{r},j}\sim \mathcal{CN} (0, \sigma_\text{r}^{2}\mathbf{I}_{M})$ denote the radar cross section (RCS) of target-$l$ and the complex AWGN at receive BS-$j$, respectively.
The steering vector $\mathbf{a} (\theta_{j,l})$ of BS-$j$'s antenna array is specifically defined as
\begin{equation}
\label{eq:steer}
\begin{aligned}
\hspace{-0.25cm}\mathbf{a} (\theta_{j,l})\triangleq \beta_{j,l}[1, e^{\jmath \frac{2 \pi}{\lambda} d \sin(\theta_{j,l})},...,e^{\jmath \frac{2 \pi}{\lambda} (M-1) d \sin(\theta_{j,l})}]^T,
\end{aligned}
\end{equation}
in which $\beta_{j,l}$, $d$, $\lambda$, and $\theta_{j,l}$ represent the distance-dependent path-loss for the BS-$j$ to target-$l$ link, antenna spacing, wavelength, and the azimuth angle of target-$l$ to BS-$j$, respectively.

Then, by aggregating all the received echo signals from BSs, the echo signal to be processed at CPU is written as
\begin{equation}
\begin{aligned}
\mathbf{y}_{\text{r}} = [\mathbf{y}_{\text{r},1}^T,...,\mathbf{y}_{\text{r},J}^T]^T = \sum \limits_{l=1}^L \widehat{\mathbf{A}}_{l}\widehat{\mathbf{w}}_\text{s}+  \widehat{\mathbf{G}}^T\widehat{\mathbf{w}}_\text{s} + \widehat{\mathbf{n}}_{\text{r}},
\end{aligned}
\end{equation}
in which for brevity we define
\begin{subequations}
\begin{align}
\mathbf{A}_{j,i,l} \triangleq& \xi_{j,i,l} \mathbf{a} (\theta_{j,l})\mathbf{a}^T (\theta_{i,l}),\\
\widetilde{\mathbf{A}}_{j,l}\triangleq& [\alpha_1\mathbf{A}_{j,1,l},...,\alpha_J\mathbf{A}_{j,J,l}],\\
\widehat{\mathbf{A}}_{l}\triangleq& [(1-\alpha_1)\widetilde{\mathbf{A}}_{1,l}^T,...,(1-\alpha_J)\widetilde{\mathbf{A}}_{J,l}^T]^T,\\
\widetilde{\mathbf{G}}_{j}\triangleq& [\alpha_1\mathbf{G}_{1 , j}^T,...,\alpha_J\mathbf{G}_{J,j}^T]^T,\\
\widehat{\mathbf{G}}\triangleq& [(1-\alpha_1)\widetilde{\mathbf{G}}_{1},...,(1-\alpha_J)\widetilde{\mathbf{G}}_{J}],\\
\widehat{\mathbf{w}}_\text{s}\triangleq& [(\mathbf{W}_1\mathbf{s}_1)^T,...,(\mathbf{W}_{J}\mathbf{s}_{J})^T]^T,\\
\widehat{\mathbf{n}}_{\text{r}}\triangleq& [(1-\alpha_1)\mathbf{n}_{\text{r},1}^T,...,(1-\alpha_J)\mathbf{n}_{\text{r},J}^T]^T.
\end{align}
\end{subequations}
After processing the received signal by receive filter $\mathbf{u}_l = [\mathbf{u}_{1,l}^T,...,\mathbf{u}_{J,l}^T]^T \in \mathbb{C}^{JM \times 1},\forall l$, the radar output signal at CPU for detecting target-$l$ is obtained as
\begin{equation}
\begin{aligned}
\mathbf{u}_l^H\mathbf{y}_\text{r} =\mathbf{u}_l^H\sum \limits_{l=1}^L \widehat{\mathbf{A}}_{l}\widehat{\mathbf{w}}_\text{s}+  \mathbf{u}_l^H\widehat{\mathbf{G}}^T\widehat{\mathbf{w}}_\text{s} + \mathbf{u}_l^H\mathbf{n}_{\text{r}}.
\end{aligned}
\end{equation}
Thus, the output sensing SINR for detecting target-$l$ is
\begin{equation}
\begin{aligned}
\mathrm{SINR}_{\text{r},l}  = \frac{\mathbf{u}_l^H\mathbf{B}_l\mathbf{u}_l}{\mathbf{u}_l^H\mathbf{C}_l \mathbf{u}_l},
\end{aligned}
\end{equation}
where for simplicity we define
\begin{subequations}
\begin{align}
\overline{\mathbf{W}} \triangleq& [\mathbf{W}_1^T,...,\mathbf{W}_J^T]^T,\; \mathbf{B}_l\triangleq \widehat{\mathbf{A}}_{l}\overline{\mathbf{W}}\overline{\mathbf{W}}^H\widehat{\mathbf{A}}_{l}^H,\\
\mathbf{C}_l\triangleq& \hspace{-0.1cm}\sum \limits_{s = 1,s \ne l}^L \hspace{-0.1cm}\widehat{\mathbf{A}}_{s}\overline{\mathbf{W}}\overline{\mathbf{W}}^H\widehat{\mathbf{A}}_{s}^H+\widehat{\mathbf{G}}^T \overline{\mathbf{W}}\overline{\mathbf{W}}^H\widehat{\mathbf{G}}^*+ \sigma_{\text{r}}^2\mathbf{Q},\\
\mathbf{Q} \triangleq& \operatorname{blkdiag}\{(1-\alpha_1)^2\mathbf{I}_{M},..., (1-\alpha_J)^2\mathbf{I}_{M}\}.
\end{align}
\end{subequations}
\vspace{-1cm}
\subsection{Problem Formulation}
In this paper, we aim to jointly design the beamforming $\overline{\mathbf{W}}$, filter $\mathbf{u}_l,\forall l$, and BS mode vector $\boldsymbol{\alpha}$ to maximize the sum of sensing SINR, while satisfying the communication SINR requirements $\gamma_k$ of each user, total transmit power budget $P_\text{max}$, as well as constraints on the numbers of transmitters and receivers.
The optimization problem is formulated as
\begin{subequations}
\label{eq:ori_pro}
\begin{align}
\max_{\overline{\mathbf{W}},\mathbf{u}_l,\forall l,\boldsymbol{\alpha}} \quad & \sum_{l= 1}^L\mathrm{SINR}_{\text{r},l} \label{eq:ori_obj}\\
\text {s.t.} \quad& \mathrm{SINR}_{\text{c},k} \ge \gamma_k,~\forall k,  \label{eq:cons_SINR}\\
& \sum_{j=1}^J \alpha_j\|\mathbf{W}_{j}\|_F^2 \le P_\text{max},\label{eq:cons_power}\\
& \|\boldsymbol{\alpha}\|_1 \ge 1,\label{eq:cons_transmitter_g_1}\\
& M\|\boldsymbol{\alpha}\|_1 \ge K,\label{eq:cons_transmitter_g_K}\\
& \|\boldsymbol{\alpha}\|_1 \le J-1,\label{eq:cons_receiver_g_1}\\
& \alpha_j \in \{0,1\}, ~\forall j,\label{eq:cons_alpha}
\end{align}
\end{subequations}
where constraints \eqref{eq:cons_transmitter_g_1}-\eqref{eq:cons_receiver_g_1} represent that there is at least one transmitter, the number of transmit antennas is greater than users, and there is at least one receiver in the network, respectively.
The non-convex problem \eqref{eq:ori_pro} is difficult to solve due to the multi-variable coupled objective function and constraints, as well as the non-smooth constraints of binary variables.
Thus, to tackle these difficulties, we divide the problem into beamforming and filter design sub-problem and BS mode selection sub-problem.
An FP-MM based beamforming and filter design algorithm and three heuristic BS mode selection methods are proposed in the following sections.
\section{Joint Transmit Beamforming and Receive Filter Design}
In this section, we develop the algorithm for solving joint beamforming and filter design problem with fixed BS mode selection vector $\boldsymbol{\alpha}$.
To be specific, the FP-MM based algorithm is adopted to solve the beamforming and filter design sub-problem in an alternating manner. 
\subsection{Receive Filter Optimization}
With given transmit beamforming $\overline{\mathbf{W}}$ and BS mode selection vector $\boldsymbol{\alpha}$,
the sub-problem for optimizing receive filter $\mathbf{u}_l,\forall l$ can be expressed as
\begin{equation}
\label{eq:pro_u}
\begin{aligned}
\max_{\mathbf{u}_l,\forall l}  \quad  \sum \limits_{l= 1}^L\frac{\mathbf{u}_l^H\mathbf{B}_l\mathbf{u}_l}{\mathbf{u}_l^H\mathbf{C}_l \mathbf{u}_l}.
\end{aligned}
\end{equation}
Since the optimization of $\mathbf{u}_l$ for detecting target-$l$ will not impact the performance of detecting other targets, problem \eqref{eq:pro_u} can be divided into $L$ sub-problems, each of which is
\begin{equation}
\label{eq:pro_u_L}
\begin{aligned}
\max_{\mathbf{u}_l}  \quad \frac{\mathbf{u}_l^H\mathbf{B}_l\mathbf{u}_l}{\mathbf{u}_l^H\mathbf{C}_l \mathbf{u}_l}.
\end{aligned}
\end{equation}
Thus, the optimal solution $\mathbf{u}_l^\star$ for maximizing this generalized Rayleigh quotient can be obtained by the eigenvector corresponding to the maximum eigenvalue of matrix $\mathbf{C}_l^{-1}\mathbf{B}_l$.

\subsection{Transmit Beamforming Optimization}
With given receive filter $\mathbf{u}_l,\forall l$ and BS mode selection vector $\boldsymbol{\alpha}$, the optimization sub-problem of transmit beamforming $\overline{\mathbf{W}}$ can be converted into a more compact form as
\begin{subequations}
\label{eq:w_ori}
\begin{align}
\max_{\widehat{\mathbf{w}}} \quad & \sum_{l= 1}^L\frac{\widehat{\mathbf{w}}^H\mathbf{D}_{l,l}\widehat{\mathbf{w}}}{\sum_{s = 1,s \ne l}^L \widehat{\mathbf{w}}^H\mathbf{D}_{l,s}\widehat{\mathbf{w}}+\widehat{\mathbf{w}}^H\mathbf{F}_l\widehat{\mathbf{w}}+c_{\text{r},l}} \label{eq:w_ori_obj}\\
\text {s.t.} \quad& \mathrm{SINR}_{\text{c},k} \ge \gamma_k,~\forall k,  \label{eq:w_ori_cons1}\\
& \widehat{\mathbf{w}}^H\boldsymbol{\Omega}\widehat{\mathbf{w}} \le  P_\text{max},
\end{align}
\end{subequations}
in which for brevity we define
\begin{subequations}
\label{eq:au}
\begin{align}
\mathbf{D}_{l,s}\triangleq& \mathbf{I}_{K+M} \otimes \widehat{\mathbf{A}}_{s}^H\mathbf{u}_l\mathbf{u}_l^H\widehat{\mathbf{A}}_{s},\\
\mathbf{F}_l\triangleq& \mathbf{I}_{K+M} \otimes \widehat{\mathbf{G}}^*\mathbf{u}_l\mathbf{u}_l^H\widehat{\mathbf{G}}^T,\\
\widehat{\mathbf{w}}\triangleq& [\overline{\mathbf{w}}_{1}^T,...,\overline{\mathbf{w}}_{K+M}^T]^T,\;
c_{\text{r},l}\triangleq\sigma_{\text{r}}^2\mathbf{u}_l^H\mathbf{Q}\mathbf{u}_l,\\
\boldsymbol{\nu} \triangleq& \mathbf{1}_{M+K}\otimes(\boldsymbol{\alpha}\otimes\mathbf{1}_M),\;
\boldsymbol{\Omega} \triangleq \operatorname{diag}\{\boldsymbol{\nu}\}.
\end{align}
\end{subequations}
In \eqref{eq:au}, $\overline{\mathbf{w}}_{k}$ denotes the $k$-th column of $\overline{\mathbf{W}}$.
It is obvious that problem \eqref{eq:w_ori} is still hard to solve because of the non-convex sum-of-ratio objective function \eqref{eq:w_ori_obj}.
Thus, we first deal with the fractional function by adopting quadratic transform and then convert the problem into a tractable form of $\widehat{\mathbf{w}}$ \cite{K. Shen}.

To be specific, by introducing an auxiliary variable $\boldsymbol{\tau}$ and removing the irrelevant constant term $c_{\text{r},l}$, the objective function \eqref{eq:w_ori_obj} can be reformulated as
\begin{equation}
\label{eq:obj_fp}
\begin{aligned}
\sum \limits_{l= 1}^L 2 \tau_l \sqrt{\widehat{\mathbf{w}}^H\mathbf{D}_{l,l}\widehat{\mathbf{w}}} - \sum \limits_{l= 1}^L\tau_l^2\widehat{\mathbf{w}}^H(\sum \limits_{s = 1,s \ne l}^L \hspace{-0.2cm} \mathbf{D}_{l,s}+\mathbf{F}_l)\widehat{\mathbf{w}},
\end{aligned}
\end{equation}
which is concave with respect to $\boldsymbol{\tau}$ and the optimal solution $\tau_l^\star$ in each iteration can be obtained by
\begin{equation}
\label{eq:tau}
\begin{aligned}
\tau_l^\star = \frac{\sqrt{\widehat{\mathbf{w}}^H\mathbf{D}_{l,l}\widehat{\mathbf{w}}}}{\sum _{s = 1,s \ne l}^L \widehat{\mathbf{w}}^H\mathbf{D}_{l,s}\widehat{\mathbf{w}}+\widehat{\mathbf{w}}^H\mathbf{F}_l\widehat{\mathbf{w}}+c_{\text{r},l}}.
\end{aligned}
\end{equation}
Unfortunately, the objective function is still difficult to solve because of the non-convex part $\sqrt{\widehat{\mathbf{w}}^H\mathbf{D}_{l,l}\widehat{\mathbf{w}}}$.
Therefore, the MM method is utilized to construct a favorable surrogate function of \eqref{eq:obj_fp} for variable $\widehat{\mathbf{w}}$ \cite{Y. Sun}.
By adopting the first-order Taylor expansion, the concave lower-bound for $\sqrt{\widehat{\mathbf{w}}^H\mathbf{D}_{l,l}\widehat{\mathbf{w}}}$ at point $\widehat{\mathbf{w}}_t$ can be derived as
\begin{equation}
\begin{aligned}
\hspace{-0.3cm}\sqrt{\widehat{\mathbf{w}}^H\mathbf{D}_{l,l}\widehat{\mathbf{w}}}\ge \sqrt{\widehat{\mathbf{w}}_t^H\mathbf{D}_{l,l}\widehat{\mathbf{w}}_t} + \Re\Big\{ \frac{\widehat{\mathbf{w}}_t^H\mathbf{D}_{l,l}(\widehat{\mathbf{w}}-\widehat{\mathbf{w}}_t)}
{\sqrt{\widehat{\mathbf{w}}_t^H\mathbf{D}_{l,l}\widehat{\mathbf{w}}_t}}\Big\}.
\end{aligned}
\end{equation}
Then, by removing the irrelevant constant terms, problem \eqref{eq:w_ori} is converted into
\begin{subequations}
\label{eq:pro_w_final}
\begin{align}
\max_{\widehat{\mathbf{w}}} \quad & \sum \limits_{l= 1}^L 2 \tau_l \Re\Big\{ \frac{\widehat{\mathbf{w}}_t^H\mathbf{D}_{l,l}\widehat{\mathbf{w}}}
{\sqrt{\widehat{\mathbf{w}}_t^H\mathbf{D}_{l,l}\widehat{\mathbf{w}}_t}}\Big\}\\ \nonumber &
- \sum \limits_{l= 1}^L\tau_l^2\widehat{\mathbf{w}}^H(\sum \limits_{s = 1,s \ne l}^L \mathbf{D}_{l,s}+\mathbf{F}_l)\widehat{\mathbf{w}} \\
\text {s.t.} \quad& \mathrm{SINR}_{\text{c},k} \ge \gamma_k,~\forall k, \\
& \widehat{\mathbf{w}}^H\boldsymbol{\Omega}\widehat{\mathbf{w}} \le P_\text{max},
\end{align}
\end{subequations}
which is a typical second-order cone programming (SOCP) problem and can be easily solved by various existing optimization algorithms or solvers.

\subsection{Summary}
The proposed FP-MM based algorithm for jointly optimizing beamforming $\widehat{\mathbf{w}}$ and filter $\mathbf{u}_l,\forall l$ is summarized in Algorithm \ref{alg:Algorithm all}.
Specifically, $\widehat{\mathbf{w}}$ and $\mathbf{u}_l,\forall l$ are updated in an alternating manner until the objective function achieves convergence.
The initial $\widehat{\mathbf{w}}$ is obtained by solving the minimum communication SINR maximization problem under the total power budget,
which can be easily solved \cite{A. Wiesel}.
Notice that this algorithm is adopted to design the beamforming and filter after obtaining the BS mode selection results by utilizing the heuristic methods proposed in the next section.
\begin{algorithm}[!t]
\begin{small}
\caption{Joint transmit beamforming and receive filter optimization algorithm.}
\label{alg:Algorithm all}
\begin{algorithmic}[1]
\REQUIRE $J$, $K$, $L$, $M$, $d$, $\lambda$, $\sigma_{\text{r}}^2$, $\sigma_{\text{c}}^2$, $\gamma_k$, $P_\text{max}$, $\boldsymbol{\alpha}$, $\mathbf{G}_{i,j}$, $\mathbf{h}_{j,k}$, $\theta_{j,l}$, and $\xi_{j,i,l}$, $\forall i,j,k$.
\ENSURE {$\widehat{\mathbf{w}}$ and $\mathbf{u}_l,\forall l$.}
\STATE {Initialize $\widehat{\mathbf{w}}$.}
\REPEAT
    \STATE{Update $\mathbf{u}_l,\forall l$ by solving problem \eqref{eq:pro_u_L};}
    \STATE{Update $\boldsymbol{\tau}$ by \eqref{eq:tau};}
    \STATE{Update $\widehat{\mathbf{w}}$ by solving problem \eqref{eq:pro_w_final};}
\UNTIL {convergence.}
\RETURN {$\widehat{\mathbf{w}}$ and $\mathbf{u}_l,\forall l$.}
\end{algorithmic}
\end{small}
\end{algorithm}

\section{BS Mode Selection}
The BS mode selection is an NP-hard problem and highly challenging to solve because of the non-smooth binary variables and coupled performance metrics.
Thus, in this section, we propose three low-complexity heuristic methods refer to as communication-centric method, sensing-centric method, and joint communication and sensing method, to efficiently obtain appropriate selection results.
\subsection{Communication-Centric}
The core idea of the communication-centric method is that the BSs which contribute less to communication are selected as receivers.
For brevity, let us first define the transmit BS and receive BS sets as $\mathcal{T}$ and $\mathcal{R}$, respectively.
All of the BSs are initialized as transmitters at the beginning, i.e., $\mathcal{T} = \{1,...,J\},~\mathcal{R}= \varnothing$.
Then, in each iteration, to find the BS in set $\mathcal{T}$ that contributes least to communication, we formulate a power minimization problem aiming to optimize the communication beamforming matrices $\mathbf{W}_{\mathrm{c}, j}, \forall j \in \mathcal{T}$ under the communication SINR requirements.
By defining the transmit power of BS-$j$ as $P_j$, the problem is written as
\begin{subequations}
\label{eq:pro_power_min}
\begin{align}
\min_{\mathbf{W}_{\mathrm{c}, j},P_j,\forall j \in \mathcal{T}}& \quad \sum_{j \in \mathcal{T}} P_j \\
\text {s.t.} \quad & \frac{ \Big| \sum_{j \in \mathcal{T}} \mathbf{h}_{j,k}^H \mathbf{w}_{j,k} \Big|^2}{\sum_{i=1,i \ne k}^{K}\Big|\sum_{j \in \mathcal{T}}\mathbf{h}_{j,k}^H \mathbf{w}_{j,i} \Big|^2 + \sigma_{\text{c},k}^{2}} \geq \gamma_k, \forall k, \\
& \left\|\mathbf{W}_{\mathrm{c}, j}\right\|_F^2 \leq P_j, \forall j \in \mathcal{T},
\end{align}
\end{subequations}
which is an SOCP problem and can be easily solved by CVX.
After solving problem \eqref{eq:pro_power_min}, the power consumptions of BSs in set $\mathcal{T}$ can be obtained.
Similar to the mechanism of water-filling algorithm, the BS having the least power consumption implies the least contribution to communications due to the worst channel condition, which should be selected as an echo signal receiver instead of transmitter.

Moreover, the sensing beamforming $\mathbf{W}_{\mathrm{r}}$ is obtained by adopting linear precoding method, which will be utilized to determine whether the BS selection process is stopped.
To be specific, we first define the steering vector of BSs for target-$l$ as $\mathbf{a}_l \triangleq [\mathbf{a}^T (\theta_{1,l}),...,\mathbf{a}^T (\theta_{J,l})]^T$, and then project the steering vector $\mathbf{a}_l$ onto the null-space of communication channels, aiming at nulling the interference of sensing to communication users.
Thus, the sensing beamforming is formulated as
\begin{subequations}
\label{eq:w_linear}
\begin{align}
\mathbf{w}_{\mathrm{au}} & =\sum_{l=1}^L \frac{(\mathbf{I}_{|\mathcal{T}| M}-\mathbf{H}(\mathbf{H}^H \mathbf{H})^{-1} \mathbf{H}^H) \mathbf{a}_l}{\left\|(\mathbf{I}_{|\mathcal{T}| M}-\mathbf{H}(\mathbf{H}^H \mathbf{H})^{-1} \mathbf{H}^H) \mathbf{a}_l\right\|_2}, \\
\mathbf{W}_{\mathrm{r}} & =\sqrt{\frac{P}{M}} \frac{\mathbf{w}_{\mathrm{au}}}{\left\|\mathbf{w}_{\mathrm{au}}\right\|_2} \otimes \mathbf{1}_M^T,
\end{align}
\end{subequations}
where $|\mathcal{T}|$ and $P$ represent the number of transmitters and power allocated for sensing, respectively.
$\otimes$ denotes the Kronecker product.
Notice that $\mathbf{W}_{\mathrm{r}}$ is stacked by sensing beamforming $\mathbf{W}_{\mathrm{r},j}$ of transmitters, where $j \in \mathcal{T}$.
Here, $P = P_{\mathrm{max}}-\sum_{j \in \mathcal{T}} P_j$ indicates the remaining power after satisfying communication requirements.

The communication-centric BS mode selection method is summarized in Algorithm \ref{alg:Algorithm 2}.
Specifically, in each iteration, the stop condition is: If the number of transmitters will not satisfy constraints \eqref{eq:cons_transmitter_g_1} and \eqref{eq:cons_transmitter_g_K} of original problem or the sum of sensing SINR decreases, the BS selection process stops.
\begin{algorithm}[!t]
\begin{small}
\caption{Communication-centric (C-C).}
\label{alg:Algorithm 2}
\begin{algorithmic}[1]
\REQUIRE $J$, $K$, $L$, $M$, $d$, $\lambda$, $\sigma_{\text{r}}^2$, $\sigma_{\text{c}}^2$, $\gamma_k$, $P_\text{max}$, $\mathbf{G}_{i,j}$, $\mathbf{h}_{j,k}$, $\theta_{j,l}$, and $\xi_{j,i,l}$, $\forall i,j,k$.
\ENSURE  {$\boldsymbol{\alpha}$, $\widehat{\mathbf{w}}$, and $\mathbf{u}_l,\forall l$.}
\STATE {Initialize $\mathcal{T} = \{1,...,J\}$, $\mathcal{R} = \varnothing$, and $\boldsymbol{\alpha} = \mathbf{1}_J$.}
\REPEAT
        \STATE{Solve problem \eqref{eq:pro_power_min};}
        \STATE{Select the receiver: $j^\star = \arg \max \limits_{j \in \mathcal{T}}\; P_j$, $\mathcal{T} \backslash j^{\star},~\mathcal{R} \cup j^{\star}$;}
        \STATE{Update the sensing beamforming by \eqref{eq:w_linear}, where $P = P_{\mathrm{max}}-\sum_{j \in \mathcal{T}} P_j$;}
        \STATE{Update the receive filter by solving problem \eqref{eq:pro_u_L};}
        \STATE{Calculate the sum of sensing SINR;}
\UNTIL{$(|\mathcal{T}|-1)M<K$ or $|\mathcal{T}| = 1$ or sum of sensing SINR decreases.}
\STATE{Re-design $\widehat{\mathbf{w}}$ and $\mathbf{u}_l,\forall l$ by Algorithm \ref{alg:Algorithm all};}
\RETURN $\boldsymbol{\alpha}$, $\widehat{\mathbf{w}}$, and $\mathbf{u}_l,\forall l$.
\end{algorithmic}
\end{small}
\end{algorithm}
\subsection{Sensing-Centric}
The core idea of the sensing-centric method is opposite of the communication-centric method.
To be specific, we still initialize the transmit BS and receive BS sets as $\mathcal{T} = \{1,...,J\},~\mathcal{R}= \varnothing$.
However, in each iteration, we aim to find the receiver that contributes more to sensing.
The contribution is calculated by assuming each BS-$j$ in set $\mathcal{T}$ is selected as a receiver, while the modes of other BSs are unchange.
The linear precoding \eqref{eq:w_linear} is adopted to obtain the transmit sensing beamforming of transmit BSs in set $\mathcal{T}\backslash j$.
Here, all the power budget is allocated for sensing, i.e., $P$ in \eqref{eq:w_linear} is set as $P_\text{max}$.
After updating the receive filter, the sum of sensing SINR of the system where BS-$j$ is set as a receiver can be calculated as $\Gamma_j = \sum \limits_{l= 1}^L\mathrm{SINR}_{\text{r},l}(\alpha_j = 0),\forall j \in \mathcal{T}$, which indicates the contribution of setting BS-$j$ as a receiver to the sensing performance.
The BS with the highest contribution will be selected as an echo signal receiver.
Then, the communication beamforming will be optimized by solving problem \eqref{eq:pro_power_min} to satisfy the communication requirements.

The sensing-centric BS mode selection method is summarized in Algorithm \ref{alg:Algorithm 3}.
Similar to the communication-centric method, the BS selection process stops when the number of transmitters will not satisfy constraints \eqref{eq:cons_transmitter_g_1} and \eqref{eq:cons_transmitter_g_K} in the next iteration or the sum of sensing SINR decreases.

\begin{algorithm}[!t]
\begin{small}
\caption{Sensing-centric (S-C).}
\label{alg:Algorithm 3}
\begin{algorithmic}[1]
\REQUIRE $J$, $K$, $L$, $M$, $d$, $\lambda$, $\sigma_{\text{r}}^2$, $\sigma_{\text{c}}^2$, $\gamma_k$, $P_\text{max}$, $\mathbf{G}_{i,j}$, $\mathbf{h}_{j,k}$, $\theta_{j,l}$, and $\xi_{j,i,l}$, $\forall i,j,k$.
\ENSURE {$\boldsymbol{\alpha}$, $\widehat{\mathbf{w}}$, and $\mathbf{u}_l,\forall l$.}
\STATE {Initialize $\mathcal{T} = \{1,...,J\}$, $\mathcal{R} = \varnothing$, and $\boldsymbol{\alpha} = \mathbf{1}_J$.}
\REPEAT
    \FOR {$j \in \mathcal{T}$}
        \STATE{Set BS-$j$ as receiver, $\alpha_j = 0$;}
        \STATE{Update the sensing beamforming by \eqref{eq:w_linear}, $P = P_{\mathrm{max}}$;}
        \STATE{Update the receive filter by solving problem \eqref{eq:pro_u_L};}
        \STATE{Calculate the sum of sensing SINR $\Gamma_j$;}
    \ENDFOR
    \STATE{$j^\star = \arg \max \limits_{j \in \mathcal{T}} \; \Gamma_j $, $\mathcal{T} \backslash j^{\star}, \mathcal{R} \cup j^{\star}$;}
    \STATE{Update communication beamforming by solving \eqref{eq:pro_power_min};}
    \STATE{Update the receive filter by solving problem \eqref{eq:pro_u_L};}
    \STATE{Calculate the sum of sensing SINR;}
\UNTIL{$(|\mathcal{T}|-1)M < K$ or $|\mathcal{T}| = 1$ or sum of sensing SINR decreases.}
\STATE{Re-design $\widehat{\mathbf{w}}$ and $\mathbf{u}_l,\forall l$ by Algorithm \ref{alg:Algorithm all};}
\RETURN $\boldsymbol{\alpha}$, $\widehat{\mathbf{w}}$, and $\mathbf{u}_l,\forall l$.
\end{algorithmic}
\end{small}
\end{algorithm}

\subsection{Joint Communication and Sensing}
The core idea of joint communication and sensing method is combining the beamforming and filter design in Algorithm \ref{alg:Algorithm all} with the BS mode selection, which jointly considers the performance of communication and sensing.
Specifically, we first initialize all BSs as transmitters and optimize beamforming $\widehat{\mathbf{w}}$. 
Then, in each iteration, before updating the receive filter $\mathbf{u}_l,\forall l$, the auxiliary variable $\boldsymbol{\tau}$, and the transmit beamforming $\widehat{\mathbf{w}}$, a receiver selection process is considered, in which the BS with highest contribution to sensing will be selected as an echo signal receiver.
Similar to the sensing-centric method, the contribution of BS-$j,\forall j \in \mathcal{T}$ as a receiver is measured by the sum of sensing SINR $\Gamma_j$.
The difference is that in the joint communication and sensing method, the beamforming for calculating $\Gamma_j$ is obtained in the previous iteration, which considers both the communication and sensing performance.
Algorithm \ref{alg:Algorithm 4} summarizes the joint communication and sensing BS selection method.
The same stop conditions are adopted.
\begin{algorithm}[!t]
\begin{small}
\caption{Joint communication and sensing (Joint).}
\label{alg:Algorithm 4}
\begin{algorithmic}[1]
\REQUIRE $J$, $K$, $L$, $M$, $d$, $\lambda$, $\sigma_{\text{r}}^2$, $\sigma_{\text{c}}^2$, $\gamma_k$, $P_\text{max}$, $\mathbf{G}_{i,j}$, $\mathbf{h}_{j,k}$, $\theta_{j,l}$, and $\xi_{j,i,l}$, $\forall i,j,k$.
\ENSURE {$\boldsymbol{\alpha}$, $\widehat{\mathbf{w}}$, and $\mathbf{u}_l,\forall l$.}
\STATE {Initialize $\mathcal{T} = \{1,...,J\}$, $\mathcal{R} = \varnothing$, $\boldsymbol{\alpha} = \mathbf{1}_J$, and $\widehat{\mathbf{w}}$.}
\REPEAT
    \FOR {$j\in\mathcal{T}$}
        \STATE{Set BS-$j$ as receiver, $\alpha_j = 0$;}
        \STATE{Update receive filter by solving problem \eqref{eq:pro_u_L};}
        \STATE{Calculate the sum of sensing SINR $\Gamma_j$ based on beamforming obtained in the previous iteration;}
    \ENDFOR
    \STATE{$j^\star = \arg \max \limits_{j \in \mathcal{T}}\; \Gamma_j$, $\mathcal{T} \backslash j^{\star}, \mathcal{R} \cup j^{\star}$;}
    \STATE{Update receive filter by solving problem \eqref{eq:pro_u_L};}
    \STATE{Update $\boldsymbol{\tau}$ by \eqref{eq:tau};}
    \STATE{Update transmit beamforming by solving problem \eqref{eq:pro_w_final};}
    \STATE{Calculate the sum of sensing SINR;}
\UNTIL{$(|\mathcal{T}|-1)M < K$ or $|\mathcal{T}| = 1$ or sum of sensing SINR decreases.}
\STATE{Re-design $\widehat{\mathbf{w}}$ and $\mathbf{u}_l,\forall l$ by Algorithm \ref{alg:Algorithm all};}
\RETURN $\boldsymbol{\alpha}$, $\widehat{\mathbf{w}}$, and $\mathbf{u}_l,\forall l$.
\end{algorithmic}
\end{small}
\end{algorithm}

\section{Simulation Results}
\begin{figure*}[t]
	\centering
	\begin{minipage}[t]{0.31\linewidth}
		\centering
		\includegraphics[width=2.6in]{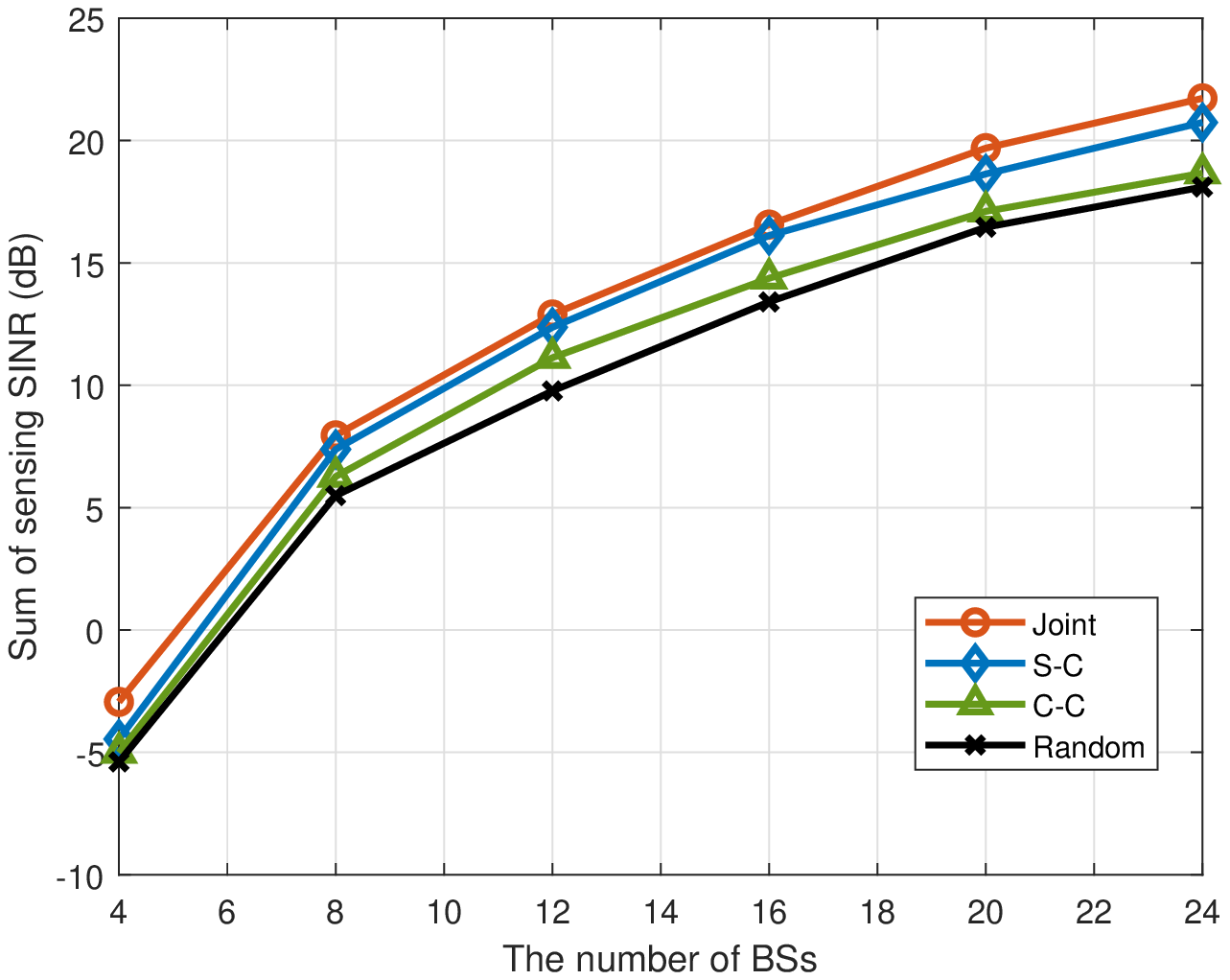}
		\caption{Sum of sensing SINR versus the number of BSs ($K = 8$, $\gamma = 8\text{dB}$, $P_\text{max}=30\text{dBm}$).}
	    \label{fig:J}
	\end{minipage}%
    \hspace{0.3cm}
	\begin{minipage}[t]{0.31\linewidth}
		\centering
		\includegraphics[width=2.6in]{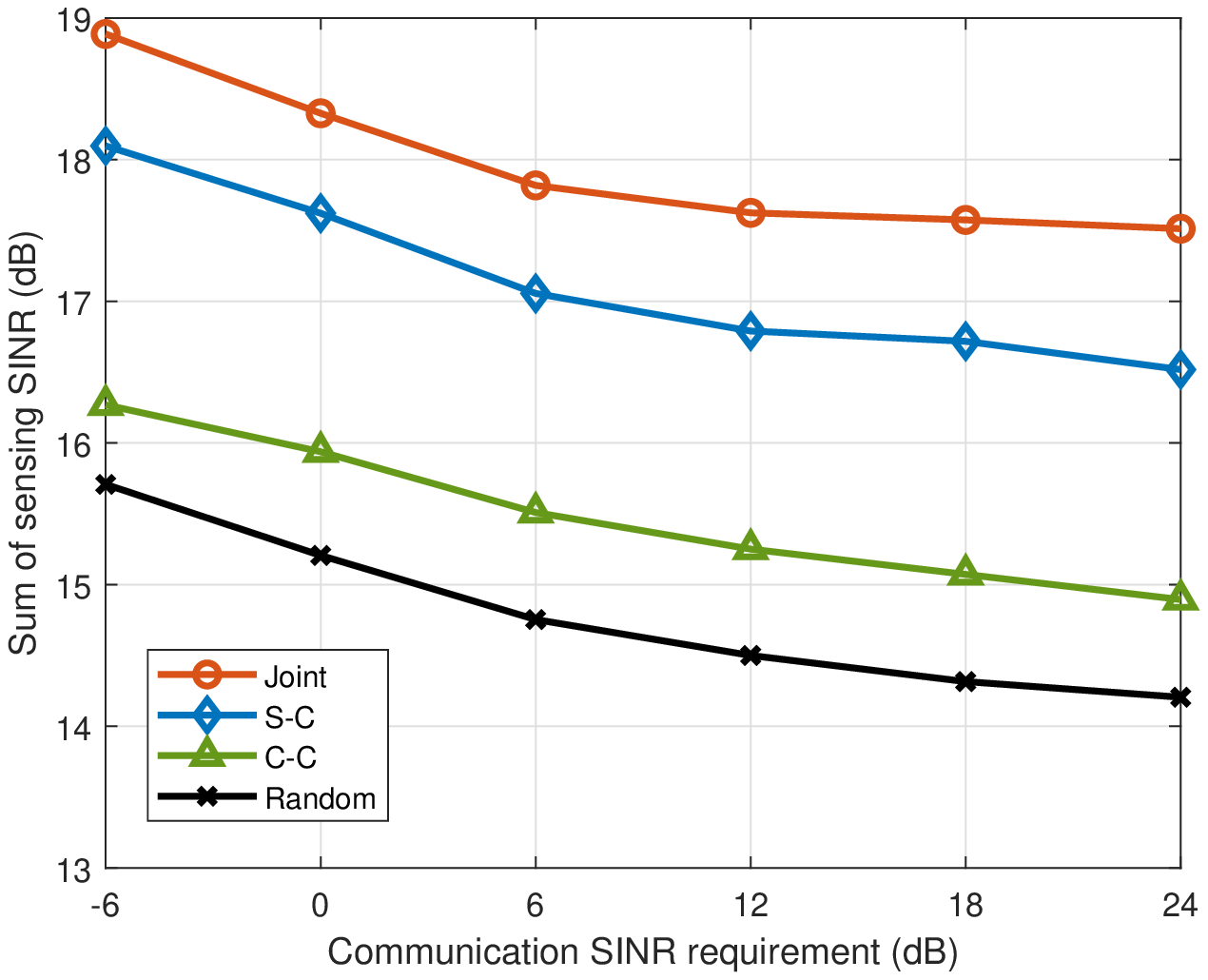}
		\caption{Sum of sensing SINR versus communication SINR requirement ($J = 16$, $K = 8$, $P_\text{max}=30\text{dBm}$).}
	    \label{fig:gamma}
	\end{minipage}
    \hspace{0.3cm}
	\begin{minipage}[t]{0.32\linewidth}
		\centering
		\includegraphics[width=2.6in]{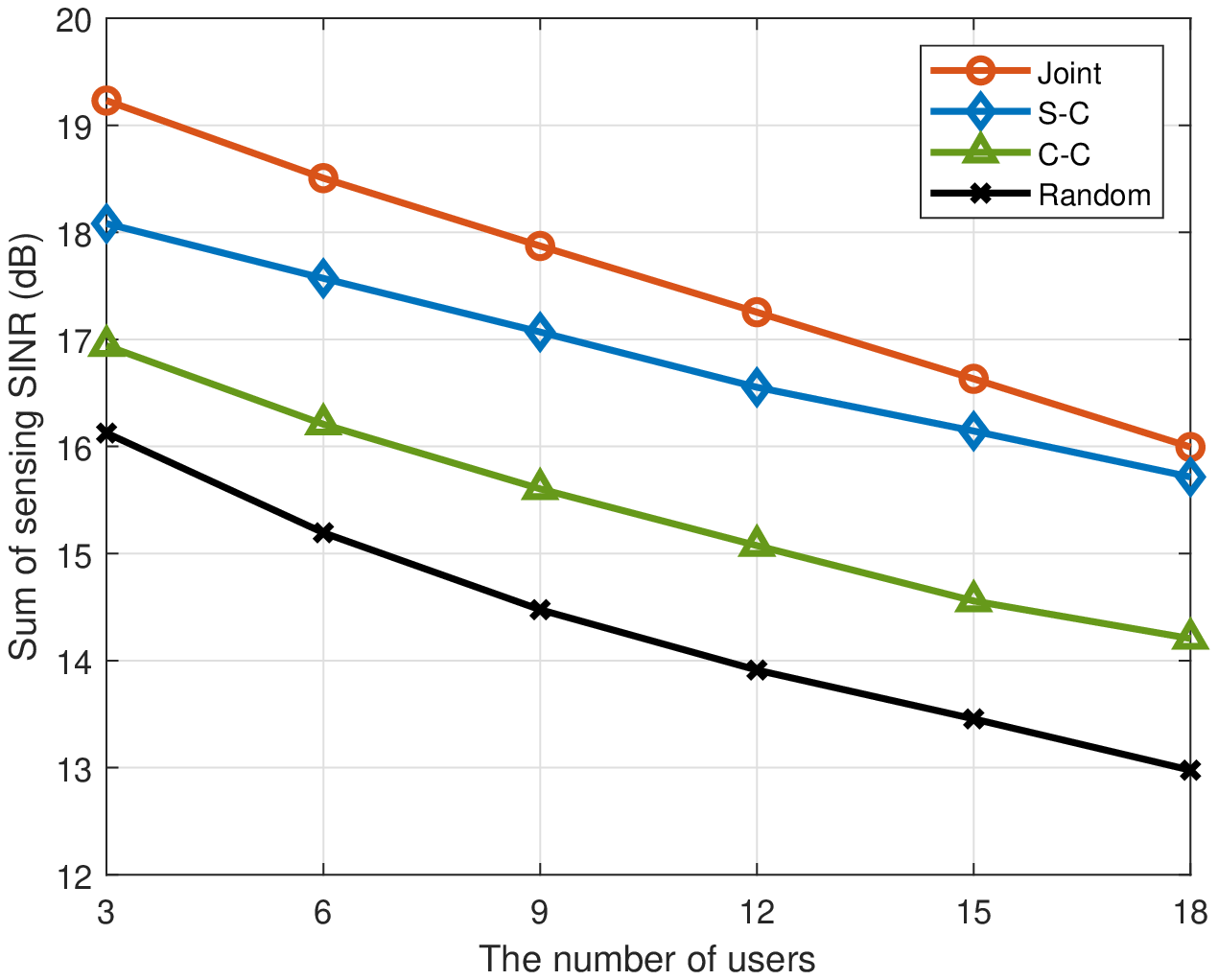}
		\caption{Sum of sensing SINR versus the number of users ($J = 16$, $\gamma = 8\text{dB}$, $P_\text{max}=30\text{dBm}$).}
	    \label{fig:K}
	\end{minipage}
\vspace{-0.3cm}
\end{figure*}
The simulation results are provided in this section to demonstrate the advancements of cooperative ISAC networks, the importance of BS mode selection, and the effectiveness of the proposed algorithms.
The performance comparison of different selection methods are also evaluated.
We consider a cooperative cell-free ISAC network consists of $J = 16$ BSs, each of which is equipped with $M = 4$ antennas and can operate as either transmitter or receiver.
The BSs cooperatively serve $K = 8$ single-antenna users and detect $L = 3$ targets.
The BSs, users, and targets are randomly located in a circle centered at $(0 \text{m}, 0 \text{m})$ with radius $D = 100\text{m}$.
The distance-dependent path-loss model is adopted, where the path-loss exponents are set as $2.2$, $2.5$, and $3.8$ for BS-target, BS-user, and BS-BS links, respectively.
The radar RCS, noise power for sensing and for communication are set as $\sigma_\text{t}^2=1$, $\sigma_\text{r}^2=\sigma_{\text{c}}^2= -80\text{dBm} $, respectively.
Moreover, the communication SINR requirements of each user are the same, i.e., $\gamma_k = \gamma, ~\forall k$.
The power budget for the network is set as $P_\text{max} = 30\text{dBm}$.
Besides, to better demonstrate the performance improvement provided by BS mode selection, the scheme with random BS mode is included in the simulations.

Fig. \ref{fig:J} evaluates the BS mode selection methods by showing the sum of sensing SINR versus the number of BSs.
It can be observed that the sensing performance increases with the increasing number of BSs even under the same total power budget, which illustrates the advantages of the cooperative cell-free ISAC networks.
Besides, the importance of the BS mode selection is verified by the fact that all of the proposed BS selection methods outperform the random selection scheme in Fig. \ref{fig:J}.
Moreover, it is obvious that the joint communication and sensing BS selection method obtains the best sensing performance, while the communication-centric method performs worst in the three proposed methods.
This is because the joint communication and sensing method considers both the communication and sensing performance in the process of BS selection, which achieves the trade-off between communication and sensing.
However, the sensing-centric and communication-centric methods only take the sensing or communication performance as primary consideration, which ignores the interaction between communication and sensing.

In Fig. \ref{fig:gamma}, the sum of sensing SINR versus the communication SINR requirement is plotted.
Similar conclusions can be drawn that the proposed BS selection approaches can achieve better performance for all communication SINR ranges.
Besides, the trade-off between communication and sensing is further presented in Fig. \ref{fig:gamma} since larger communication SINR requires more resources allocated to communication users, which results in the decreasing in sensing performance.

Finally, the sum of sensing SINR versus the number of users is illustrated in Fig. \ref{fig:K}. Not surprisingly, the sensing performance decreases with increasing number of users since more resources are allocated to communications. Moreover, the proposed mode selection methods perform better than the random method, which further verifies the importance of BS mode selection and effectiveness of proposed methods.

\section{Conclusion}
In this paper, we investigated the joint design of BS mode selection, transmit beamforming, and receiver filters for cooperative cell-free ISAC networks.
An efficient FP-MM based joint beamforming and filter design algorithm and three low-complexity heuristic BS mode selection methods were proposed to solve the sum of sensing SINR maximization problem under the communication SINR requirements, total power budget, and constraints on the numbers of transmitters and receivers.
Simulation results verified the significant performance improvement provided by multi-BS cooperation and BS mode selection in cooperative cell-free ISAC networks.
The effectiveness of the proposed joint design algorithms is also illustrated.

\end{document}